\documentclass{article} 
\usepackage{iclr2015,times,graphicx}
\usepackage{hyperref}
\usepackage{url}

\title{Audio Source Separation Using a Deep Autoencoder}

\author{
Giljin Jang	\\
School of Electronics Engineering	\\
Kyungpook National University	\\
Daegu, Republic of Korea	\\
\texttt{gjang@ee.knu.ac.kr}	\\
\AND
Han-Gyu Kim \& Yung-Hwan Oh	\\
Computer Science Department	\\
Korea Advanced Institute of Science and Technology (KAIST)	\\
Daejeon, Republic of Korea	\\
\texttt{\{hgkim,yhoh\}@cs.kaist.ac.kr}
}

\begin{document}

\maketitle

\begin{abstract}
This paper proposes a novel framework for unsupervised audio source separation using a deep autoencoder.
The characteristics of unknown source signals mixed in the mixed input is automatically by properly configured autoencoders implemented by a network with many layers, and separated by clustering the coefficient vectors in the code layer.
By investigating the weight vectors to the final target, representation layer, the primitive components of the audio signals in the frequency domain are observed. By clustering the activation coefficients in the code layer, the previously unknown source signals are segregated.
The original source sounds are then separated and reconstructed by using code vectors which belong to different clusters.
The restored sounds are not perfect but yield promising results for the possibility in the success of many practical applications.
\end{abstract}

\section{Introduction}
\label{sec:introduction}

In audio analysis applications such as speech recognition and audio-text alignment, the clean target sound helps improve the quality of analysis while noisy input may disturb the analysis process.
In a situation that we only have recordings from a single sensor, the problem is very difficult and hence no general solution has been found.
Masking in the spectro-temporal region was proposed in ~\citet{HuWang04} where the separation mask was constructed by estimated speech pitch. Such method worked well in extracting speech from noisy environment, but the performance is not guaranteed if the target source is not a speech signal.
Several separation methods based on non-negative matrix factorization (NMF) was proposed to solve the monaural source separation problem~\citep{Virtanen10}. NMF utilized the redundancy of the sound spectrogram for source separation. NMF-based methods works for various types of sources.
However, the mixing is assumed to be a non-negative linear mixing, and it cannot handle complex sound sources.

In this paper, we propose applying a deep autoencoder~\citep{hinton2006reducing} to the single-channel, source separation problem.
The autoencoders constructed by networks with many hidden layers have been successfully adopted in many applications, such as image and audio denoising, speech recognition, data compression, and so on. Unlike other applications, our proposed method tries to apply the autoencoder to solve the problem of unsupervised audio source separation.
The characteristics of sound sources in the input mixture are learned and stored in a deep autoencoder, and an appropriate source separation algorithm based on unsupervised learning is proposed. Experimental results on mixtures of 5 types of music and 2 types of speech signals suggest that the coefficients in the encoding layer is useful in distinguishing and separating the unknown target sources.
The detailed descriptions	 are in the following sections.

\section{Source separation using autoencoder}
\label{sec:proposed}

In order to learn an autoencoder for audio source mixtures, we first apply short-time Fourier analysis to the time-domain audio signal, resulting magnitude spectrum matrix denoted by $\mathbf{X}_{c,m}$, where $c$ is the frequency channel index and $m$ the temporal frame index.
For the input to the autoencoder, we construct a rectangular window with consecutive frequency channel and time index, such that
\begin{equation}
\label{eq:windowing}
\mathbf{W}_{i,j} = \{\mathbf{X}_{c,m} | i\leq{}c<i+h, j\leq{}m<j+l \},
\end{equation}
where $\mathbf{W}_{i,j}$ is the window starting from frequency channel $i$ and frame $j$ of the magnitude spectra, where $h$ and $l$ are the height and length of the windows, respectively.
The rectangular windows are then unrolled into supervectors for the autoencoder training.
The data set is generated by shifting the window one frame in time axis or one frame in frequency axis from all the input spectrum matrix.

\begin{figure}[htb]
\begin{center}
\includegraphics[width=0.4\columnwidth]{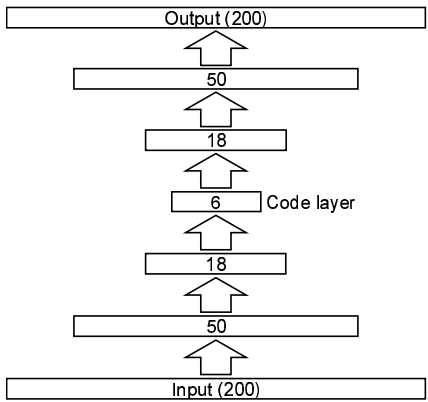}
\end{center}
\caption{The structure of autoencoder for source separation.}
\label{fig:proposed_autoencoder}
\end{figure}

We designed an autoencoder with $5$ hidden layers, which have $50$, $18$, $6$, $18$, and $50$ units, respectively, as shown in Figure~\ref{fig:proposed_autoencoder}.

\begin{figure}[htb]
\begin{center}
\includegraphics[width=0.7\columnwidth]{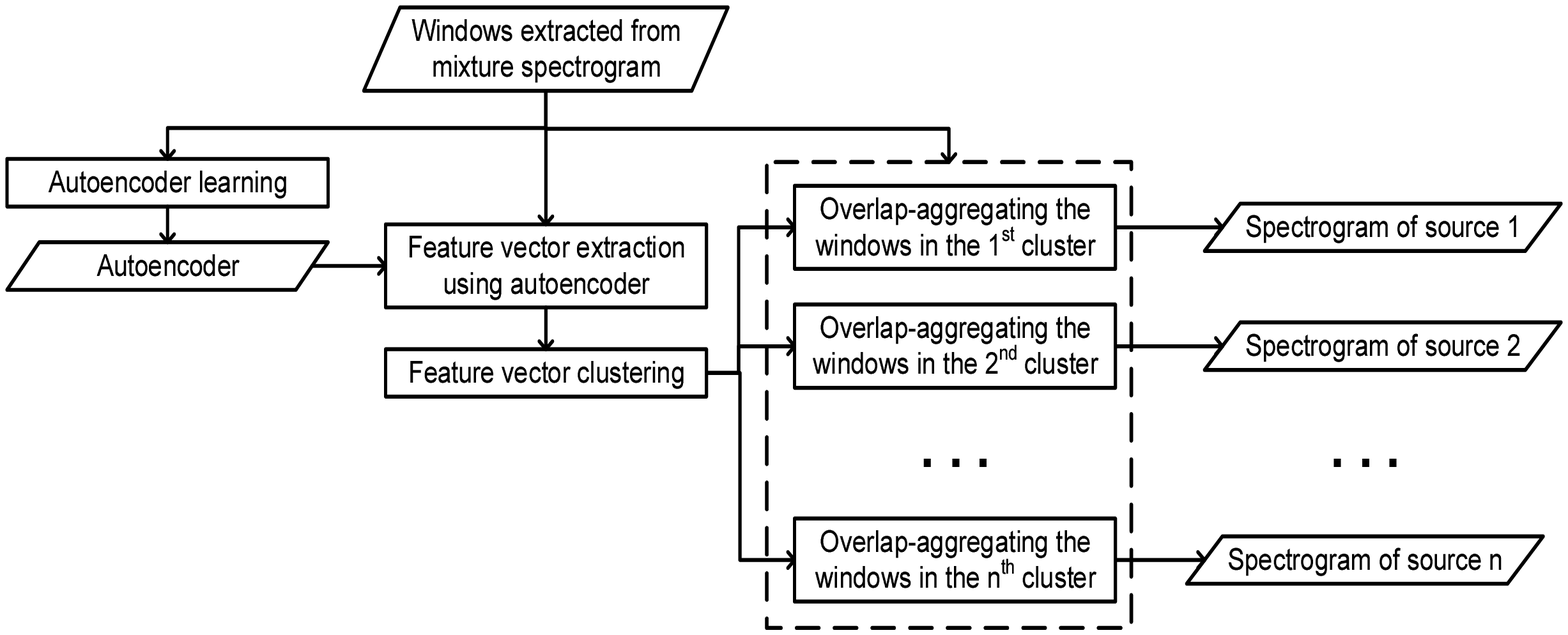}
\end{center}
\caption{The block diagram of the clustering-based source separation method.}
\label{fig:clustering_method}
\end{figure}

Because there is no labeled training data for the unknown sources, a k-means clustering is performed on the 6-dimensional coefficient vectors of the middle layer. The whole process of the clustering is shown in Figure~\ref{fig:clustering_method}.
In this method, we consider the feature vectors of the windows extracted using the autoencoder from different clusters for different sound sources. According to the clustering result of the feature vectors, the windows are also classified into the clusters, and original audio sources are reconstructed using the vectors in the given cluster only.

\section{Experimental results}
\label{sec:result}

\begin{figure}[htb]
\begin{center}
\includegraphics[width=0.4\columnwidth]{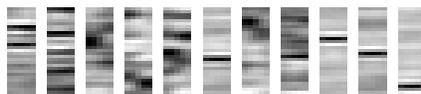}
\end{center}
\caption{Example windows constructed by the weights to the last target layer for jazz and male speech mixture.}
\label{fig:basis_example}
\end{figure}

\begin{figure}[htb]
\begin{center}
\includegraphics[width=\columnwidth]{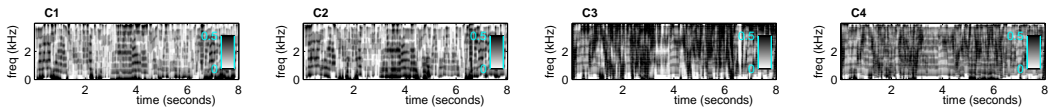}
\end{center}
\caption{Separation masks constructed by reconstructed vectors in each of 4 clusters for jazz and male speech mixture.}
\label{fig:mask_example}
\end{figure}

To evaluate the performance of the proposed method, source separation experiments were carried out on a mixtures of 5 different music sounds and 2 different speech sources. The speech sources are selected from TIMIT speech corpus, and music sounds are jazz, drum, acoustic guitar, electric guitar and piano. These speech and music sounds were mixed together, generating $2 \times 5 = 10$ mixtures in total.
The audio sounds are sampled at $8$~kHz, and only 8 seconds are used in training the respective autoencoders.
The spectrogram matrix of the mixtures are obtained by short-time Fourier analysis with frame length $40$~ms and shift size $10$~ms.
For window generation in Equation~\ref{eq:windowing}, $h = 30$ and $l = 5$ were used, which were decided empirically.
Figure~\ref{fig:basis_example} shows 10 selected examples from the total 50 weight vectors connecting all the nodes in the last hidden layer to one of the nodes in the final output layer. The input is male speech and jazz music mixture. 
Four of them represents the change of the spectral peaks over time: third, fourth, fifth, and seventh, which reflect the temporal change of harmonics of speech signals.
The remaining six vectors are composed of straight lines over time, modeling the stationary frequency components of music signals.
Figure~\ref{fig:mask_example} represents the mask constructed by clustering all the code vectors of dimension 6 in the middle layer. The first and the second clusters are jazz music, with slowly changing spectral peaks, and the third and the fourth clusters are mostly speech sounds.
These results show that the validity of the proposed method.

\section{Conclusions}
\label{sec:conclusion}

An application of autoencoders to audio source separation is proposed.
The spectrum matrix obtained by short-time Fourier analysis is used for the initial representation of the mixed source signal, and a deep autoencoder is used to represent the spatio-temporal local parts of the spectrum matrix.
The code coefficients in the middle layer of the autoencoder is used as a feature for distinguishing audio sources.
The main contribution of the proposed method is that the characteristics of unknown sources are extracted from the mixed signals.
Although the experimental results are not complete yet, ongoing efforts are being made to improve the proposed method.

\subsubsection*{Acknowledgments}

%
This work was supported by the Basic Science Research Program through the National Research Foundation of Korea (NRF) funded by the Ministry of Education, Science and Technology (no. NRF-2010-0025642).

\bibliographystyle{iclr2015}

\end{document}